\documentclass{article}
\usepackage{amssymb,amsmath,amsthm,graphicx,epsfig,latexsym}

\newcommand{\ket}[1]{|#1\rangle}
\newcommand{\bra}[1]{\langle #1 |}
\newcommand{\braket}[2]{\langle #1 | #2 \rangle}
\newcommand{\m}[1]{\overline{#1}}
\newcommand{\be}{\begin{equation*}} \newcommand{\ee}{\end{equation*}}
\newcommand{\bea}{\begin{eqnarray*}} \newcommand{\eea}{\end{eqnarray*}}

\font\german=eufm10 at 10pt
\def\Buchstabe#1{{\hbox{\german #1}}}
\def\EA{\Buchstabe{A}}      % event algebra

\def\Z2{\mathbb{Z}_2}
\newtheorem{lemma}{Lemma}

\newtheorem{claim}{Claim}

 \newcommand{\RR}{\mathbb{R}}

\begin{document}

\makeatletter
\renewcommand{\theequation}{\thesection.\arabic{equation}}
\@addtoreset{equation}{section} \makeatother

\baselineskip 18pt

%%%%%%%%%%%%%%%% title page %%%%%%%%%%%%%%%%%%%%%%%%%%%%%%%%%%%%

\begin{titlepage}

\vfill

\begin{center}
\baselineskip=16pt {\Large\bf {Dynamical Wave Function Collapse Models
in Quantum Measure Theory}}
% \vskip 5 mm Running head: Lattice collapse model\\
\vskip 10.mm Fay Dowker and Yousef Ghazi-Tabatabai\
\end{center}
%\vfill
\vfill \par
\begin{center}
{\bf Abstract}

\end{center}
\begin{quote}

The structure of Collapse Models is investigated in the framework of
Quantum Measure Theory, a histories-based approach to quantum mechanics.
The underlying structure of coupled classical and quantum systems
is elucidated in this approach which puts both systems
on a spacetime footing. The nature of the coupling is exposed:
the classical histories have no dynamics of their own but are simply
tied, more or less closely, to the quantum histories.

\end{quote}

\vfill

\hrule width 5.cm \vskip 5mm {\small \noindent Blackett
Laboratory, Imperial College, London SW7 2AZ,
UK.
\\ }
\end{titlepage}
%%%%%%%%%%%%%%%%%%%%%%%%%%%%%%%%%%%%%%%%
\setcounter{equation}{0}

\section{Introduction}

Models of ``spontaneous localisation'' or
``dynamical wavefunction collapse'' are observer independent
alternatives to standard Copenhagen quantum theory (see \cite{Bassi:2003gd}
for a review). These models have a generic
structure: there is a quantum state $\Psi$ which undergoes a stochastic
evolution in Hilbert space and there is a ``classical'' (c-number)
entity -- call it $\alpha$ -- with a stochastic evolution in spacetime.
The stochastic dynamics for the two entities -- $\Psi$ and $\alpha$ --
are coupled together. The stochastic dynamics in Hilbert space
tends to drive $\Psi$ into an eigenstate of an operator $\hat{\alpha}$
 that
corresponds to $\alpha$.
And the probability distribution for the realised values
of $\alpha$ depends on $\Psi$ so that the history of
$\alpha$ follows, noisily,
the expectation value of $\hat\alpha$ in $\Psi$.

That collapse models have both quantum and classical
aspects has been pointed out before, notably by
Di\'osi.
The nature of this interaction between the classical and
quantum parts of these models is, however, somewhat obscured
by the profound difference in the nature of their descriptions: the
classical variable traces out a history in spacetime and the
quantum state traces out its evolution in Hilbert space.

In order to illuminate the nature of the quantum-classical
coupling within collapse models we will, in the case of
a concrete and specific example, recast the formalism into
the framework of {\it {generalised measure theory}} \cite{Sorkin:1994dt}
in which both classical and quantum systems are treated
on as equal a footing as possible. The
classical variables will continue to have a spacetime
description but the quantum system will now also be
described in terms of its spacetime histories and not
fundamentally in terms of any state in Hilbert space.

The model we will focus on is a discrete, finite, 1+1 dimensional
lattice field theory. This is a useful model because it
is completely finite (so long as we restrict ourselves
to questions involving finite times)
and expressions can be written down
exactly and also because there is a well-defined
background with non-trivial causal structure, so that
questions of causality can be explored.

We will show that the model contains both ``classical''
and ``quantum'' histories, and demonstrate the nature
of their interaction. We will show that one  choice of
ontology for collapse models, the Bell ontology \cite{Bell:1987i},
corresponds to coarse graining
over the quantum histories. We will also show how the
well-known relationship between collapse models and open
quantum systems coupled to an environment reveals itself
in this histories framework.

\section{Quantum measure theory}\label{sec:QMT}

We start with a brief review of generalised
measure theory and quantum
measure theory and refer to \cite{Sorkin:1994dt,Sorkin:1995nj,Salgado:1999pu,
Sorkin:2006wq,Sorkin:2007}
for more details.

A generalized measure theory consists of a triple, $(\Omega, \EA, \mu)$,
of a space of histories, an event algebra and a measure. The space of
histories, $\Omega$, contains all the ``fine grained histories'' or
``formal trajectories'' for the system
 {\it e.g.} for $n$-particle mechanics -- classical or quantum -- a history would be
a set of $n$ trajectories in spacetime,
and for a scalar field theory, a history
would be a field configuration on spacetime.

The event algebra, $\EA$, contains all the (unasserted) propositions that
can be made about the system. We will call elements of
$\EA$ {\textit{ events}}, following standard terminology in the theory
of stochastic processes.   In cases where $\Omega$ is finite,
$\EA$ can be identified with the power set, $2^\Omega$. When $\Omega$ is
infinite, $\EA$ can be identified with an appropriate ring of sets
contained in the
power set: $\EA \subset 2^\Omega$.\footnote {$\EA$ is a Boolean algebra
with addition in the algebra corresponding to
 symmetric difference and multiplication in the algebra corresponding
to intersection. We will not employ this algebraic notation in this
paper.}

Predictions about the system --- the dynamical
content of the theory ---  are to be gleaned,
in some way or another, from
a generalized measure $\mu$, a non-negative real function on
$\EA$. $\mu$ is the dynamical law and initial condition rolled into
one.

Given the measure, we can construct the following series of symmetric set functions,which are sometimes referred to as the Sorkin hierarchy\footnote{These are the generalised interference terms introduced in \cite{Sorkin:1994dt}}:
\begin{align*}
        I_1(X) & \equiv \mu(X) \\
      I_2(X,Y) & \equiv \mu(X\sqcup Y) - \mu(X) - \mu(Y) \\
  I_3(X, Y, Z) & \equiv \mu(X\sqcup Y \sqcup Z) -
      \mu(X\sqcup Y) - \mu(Y\sqcup Z) - \mu(Z\sqcup X) \\
{}&\ + \mu(X) + \mu(Y) + \mu(Z)
\end{align*}
and so on, where $X$, $Y$, $Z$, {\it etc.\ }are disjoint elements of $\EA$,
as indicated by the symbol `$\sqcup$' for disjoint union.

A {\it measure theory of level $k$} is defined as
one which satisfies the sum rule
$I_{k+1}=0$.
It is known that this condition implies that all higher sum rules are
automatically satisfied, {\it viz.} $I_{k+n}=0$ for all $n\geq 1$
\cite{Sorkin:1994dt}.

A level 1 theory is thus
one in which the measure satisfies the usual
Kolmogorov sum rules of classical probability theory,
classical
Brownian motion
being a good example.
A level 2 theory is one in which the Kolmogorov sum rules
may be violated but $I_3$ is nevertheless zero. Any unitary quantum
theory
can be cast into the form of a generalised measure theory and
its measure
satisfies the condition $I_3 = 0$.
We refer to level 2 theories, therefore, as {\it quantum} measure theories.

The existence of a
quantum measure, $\mu$, is more or less equivalent \cite{Sorkin:1994dt} to the
existence of a
{\it decoherence functional},
$D(\,\cdot\,\,;\,\cdot\,)$, a complex function on $\EA\times \EA$
satisfying \cite{Hartle:1989,Hartle:1992as}:

\noindent (i) Hermiticity: $D(X\,;Y) = D(Y\,;X)^*$ ,\  $\forall X, Y \in \EA$;

\noindent (ii) Additivity: $D(X\sqcup Y\,; Z) = D(X\,;Z) + D(Y\,;Z)$ ,\
   $\forall X, Y, Z \in \EA$ with $X$ and $Y$ disjoint;

\noindent (iii) Positivity: $D(X\,;X)\ge0$ ,\  $\forall X\in \EA$;
% \noindent (iii) (Weak) positivity: $D(X\,;X)\ge0$ ,\  $\forall X\in \EA$;

\noindent (iv) Normalization: $D(\Omega \,;\Omega)=1$.\footnote{The normalisation
condition may turn out not to be necessary, but we include it
because all the
quantum measures we consider in this paper will satisfy it.}

\noindent The quantal measure is related to
the decoherence functional by
\begin{equation}
\mu(X) = D(X \,; X)\quad \forall X \in \EA \,.
\end{equation}
The quantity $D(X\,;Y)$ is
interpretable as the quantum interference between two sets of
histories in the case when $X$ and $Y$ are disjoint.

\section{The lattice field model}

We review the lattice field model \cite{Dowker:2002wm,
Dowker:2004zn}
whose structure we will investigate.  The model is based on a
unitary QFT on a 1+1 null lattice \cite{Destri:1987ze}, which becomes a collapse
model on the introduction of local ``hits'' driving
the state into field eigenstates.

\begin{figure}[thb]
\epsfxsize=10cm \centerline{\epsfbox{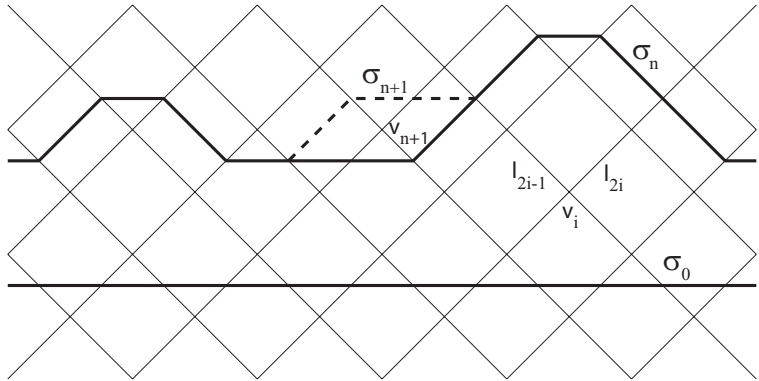}}
\caption{The light cone lattice. $\sigma_0$ is the initial
surface and $\sigma_n$ is a generic spacelike surface. The surface
$\sigma_{n+1}$ is shown after the vertex $v_{n+1}$ is evolved over.
A vertex $v_i$ is shown with its two outgoing links: $l_{2i-1}$ to the
left and $l_{2i}$ to the right.}
\label{fig:lattice}
\end{figure}

The spacetime lattice is a lightcone discretisation of a cylinder, $N$ vertices wide and periodic in space. It extends to the infinite future, and the links between the lattice vertices are left or right going null rays. Figure \ref{fig:lattice} shows a part of such a spacetime lattice, identifying the leftmost vertices with the rightmost vertices we see that $N=6$. A spacelike surface $\sigma$ is maximal set of mutually spacelike links, and consists of $N$ leftgoing links and $N$ rightgoing links cut by the surface; an example of a spatial surface is shown in figure \ref{fig:lattice}. We assume an initial spacelike surface $\sigma_0$.

An assignment of labels, $v_1, v_2, v_3, \dots$,
to the vertices to the future of $\sigma_0$
 is called ``natural'' if $i< j$ whenever the vertex labelled $v_i$ is to the
causal past of the vertex labelled $v_j$. A natural labelling is 
equivalent to a linear
extension of the (partial) causal order of the vertices. A natural
labelling, $v_1, v_2, \dots$ is also equivalent to a sequence of
spatial  surfaces, $\sigma_1, \sigma_2, \dots$ where the surface
$\sigma_n$ is defined such that between it and $\sigma_0$, lie
exactly the vertices $v_1, \dots v_n$. One can think of the
natural labelling as giving an ``evolution'' rule for the
spacelike surfaces: at time step $n$ the surface
creeps forward by one ``elementary motion'' across vertex $v_n$.
For the purpose of this paper, it is convenient to
consider a fixed natural labelling.
Nothing will depend on the natural labelling chosen, all mathematical
quantities will be independent of the choice.

The local field variables $\Phi$ live on the links. These field
variables take only two values $\{0,1\}$, so that on each link
there is a qbit Hilbert space spanned by the two field eigenstates
$\{\ket{0}, \ket{1}\}$.
As the field variables live on the links, it is convenient to
have a labelling of the links. We choose a labelling $l_a$,
$a = 1,2,\dots$, such that $l_{2i-1}$ and $l_{2i}$ are the
left-going and right-going outgoing links, respectively, from vertex $v_i$ (see figure \ref{fig:lattice}).
So, as vertex label $i$ increases from 1 to $n$, the link
label $a$ runs from $1$ to $2n$. We denote the qbit Hilbert space related to link $l_a$ by $H_{l_a}$.

The initial state $|\psi_0\rangle$ on surface $\sigma_0$ is
an element of the $2^{2N}$ dimensional Hilbert space $H_{\sigma_0}$
which is a tensor
product of the $2N$ 2-dimensional
Hilbert spaces on each link cut by $\sigma_0$, $H_{\sigma_0}=\displaystyle{\bigotimes_{l_a\in\sigma_0}}H_{l_a}$.
Similarly there is a $2^{2N}$ dimensional Hilbert space for each spacelike
surface $\sigma_i$ and they are isomorphic via the
isomorphisms, tied to the lattice, which map each link's
qbit Hilbert space onto the Hilbert spaces for the links
vertically above it on the lattice. In this way we can identify the Hilbert spaces $H_{\sigma_i}$ ($=\displaystyle{\bigotimes_{l_a\in\sigma_i}}H_{l_a}$) on each surface and describe the time evolution with a state evolving in a single Hilbert space $H_q$ ($\simeq H_{\sigma_i}$) of the system.

\subsection{The unitary theory}

In the standard unitary version of this local field theory, there
is a local unitary evolution operator, $R_i$, for each
$v_i$, which acts unitarily on the 4-dimensional factor of the Hilbert space
associated to the two ingoing and two outgoing links for $v_i$,
and acts as the identity operator on all other factors.
The state vector is evolved from $\sigma_{i-1}$
to $\sigma_i$ by applying $R_i$ \cite{Destri:1987ze}. 

So in figure \ref{fig:lattice} we see that the surface $\sigma_n$ 
evolves `over' vertex $v_{n+1}$ to give us surface $\sigma_{n+1}$. 
Now if $l_j,l_k$ are the two links going `into' vertex $v_{n+1}$, 
and $l_{2(n+1)-1},l_{2(n+1)}$ the two outgoing links, the operator 
$R_{n+1}$ maps $H_{l_j}\otimes H_{l_k}$ to 
$H_{l_{2(n+1)-1}}\otimes H_{l_{2(n+1)}}$. 
Further, for the links in the intersection of 
$\sigma_n$ and $\sigma_{n+1}$, $R_{n+1}$ acts as the the identity. 
Since the surfaces $\sigma_n, \sigma_{n+1}$ only differ on the 
links $l_j,l_k,l_{2(n+1)-1},l_{2(n+1)}$, we can put this 
together to get $R_{n+1}:H_{\sigma_n}\rightarrow H_{\sigma_{n+1}}$.

Since we have identified the Hilbert spaces $H_{\sigma_i}$, we  
regard $R_{n+1}$ as evolving a state in the `system Hilbert space' 
$H_q$, so we write:
\begin{eqnarray}
\ket{\psi_{n+1}}&=&R_{n+1}\ket{\psi_n}\nonumber \\
&=& R_{n+1} R_n \ldots R_1 \ket{\psi_0}\,. \label{eqn:quantum evolution}
\end{eqnarray}
We define the unitary time evolution operator, $U(n)$, by 
\begin{equation}\label{eq:Uoperator}
U(n) \equiv R_n\; R_{n-1} \ldots R_1\;.
\end{equation} 

To cast the theory into a quantum measure theory framework,
we need to identify the space, $\Omega_q$ of histories, an event
algebra, $\EA_q$, of suitable subsets of $\Omega_q$ and the
decoherence functional, $D_q(\,\cdot\,\,;\,\cdot\,)$.

In the lattice field theory the set of histories, $\Omega_q$, is the set 
of all field configurations on the lattice to the future of $\sigma_0$. 
A field configuration, $\Phi$, is an assignment of $0$ or $1$ to every link,
in other words $\Phi$ is a function from the infinite set of links, 
$\{l_a: a = 1,2,\dots\}$,
to $\mathbb{Z}_2$.

The  events that we want to consider are those which refer to properties
of the histories which are bounded in time. In other words
for $A \subset \Omega_q$ to be an event there must exist an integer $m$
such that to determine whether or not a field
configuration, $\Phi$ is in 
$A$ it is only necessary to
know the values of $\Phi$ between $\sigma_0$ and $\sigma_m$.
For example, the subset 
\begin{equation*}
E_k =\{\Phi \in \Omega_q: \Phi(l_{2k}) = 1\}
\end{equation*}
is an event for any fixed $k$. But the subset
\begin{equation*}
E = \{\Phi \in \Omega_q: \exists k \ \text{s.t.} \ \Phi(l_{2k}) = 1 \}
\end{equation*}
is {\it not} an event  (at least not for the purposes of the current paper).

We want to consider all events that are bounded in time. To this
end, for each positive integer $n$ we define $\Omega_q^n$ to be the 
set of field
configurations, $\Phi^n$, on the 
first $2n$ links, $l_1,\dots l_{2n}$,
outgoing from the first $n$ vertices $v_1,\dots v_n$. (Recall that 
we have chosen an arbitrary, but fixed, natural labelling of the 
vertices which gives unambiguous meaning to ``the first $2n$ links''.)
We define the cylinder set $Cyl(\Phi^n)$
to be the set of all elements of $\Omega_q$ which coincide with
$\Phi^n$ on $l_1,\dots l_{2n}$:
\begin{equation*}
Cyl(\Phi^n) \equiv \{\Phi \in \Omega_q | \Phi = \Phi^n\ {\text{when restricted
to the first}} \ 2n \ {\text{links}} \}\;.
\end{equation*}

Each cylinder set, $Cyl(\Phi^n)$
is an event that is bounded in time: it is the 
event ``the first $2n$ values of the field agree with  
$\Phi^n$.''  
The event algebra, $\EA_q$, then, is the (unital)
ring of sets generated, under finite union and intersection,
by all the cylinder sets, $Cyl(\Phi^n)$, for all $n$ and all 
$\Phi^n \in \Omega^n_q$.

Two cylinder sets have nonempty intersection if and only if one
contains the other and the complement of a cylinder set (that
for $\Phi^n$, say) is
a disjoint union of finitely many cylinder sets (those
for all the
configurations on $l_1 \dots l_{2n}$ that are {\textit{not}} $\Phi^n$).
Thus, all elements of $\EA_q$ are finite, disjoint unions of cylinder sets.
Given an event, $A \in \EA_q$,
there is indeed an integer, $m$, such that to determine whether or not a field
configuration, $\Phi$ is in $A$ it is only necessary to
know the values of $\Phi$ between $\sigma_0$ and $\sigma_m$.
We will refer to the minimum such $m$ as the {\it {time extent}}
of $A$. 
The time extent of the cylinder set $Cyl(\Phi^n)$ is clearly
$n$ and the time extent of an event $A$ is no greater than the 
maximum of the time extents of the cylinder sets whose union $A$ is. 

Consider the example given previously, $E_k$. We can see that 
this is the union of all the cylinder sets for the $\Phi^k$ such that 
$\Phi^k(l_{2k}) = 1$:
\begin{equation}
E_k = 
\mathop{\bigcup_{\Phi^k\ {\text s.t.}} }_{\Phi^k(l_{2k}) =1}
Cyl(\Phi^k)\;.
\end{equation}
The time extent of event $E_k$ is $k$.

A cylinder set is an event which corresponds to the  
history of the field up to a finite time. 
For each cylinder set, $Cyl(\Phi^n)$,
the {\textit {class}} operator, $C(\Phi^n)$ \cite{Hartle:1992as},
for that finite history is given by 
\begin{equation}\label{eq:classoperator}
C(\Phi^n) \equiv P^H_{2n}({\Phi^n_{2n}})\;
P^H_{2n-1}({\Phi^n_{2n-1}}) \dots P^H_2({\Phi^n_{2}})\;
P^H_1({\Phi^n_{1}}) \; .
\end{equation}
$P^H_a({\Phi^n_a})$ is the projection operator onto the eigenspace
corresponding to the value, $\Phi_a^n = 0$ or $1$, of $\Phi^n$  at link $l_a$,
in the Heisenberg Picture:
\begin{equation}
P^H_a(\Phi^n_a) = U([(a+1)/2])^\dagger\; P_a(\Phi^n_a)\; U([(a+1)/2])
\end{equation}
where $P_a(\Phi^n_a)$ is the Schr\"odinger Picture projector,
$U(k)$ is the unitary time evolution operator (\ref{eq:Uoperator})
and 
$[\cdot]$ denotes integer part. The Schr\"odinger picture projector is
\begin{equation}
P_a(\Phi^n_a) =
 \ket{\Phi^n_a} \bra{\Phi^n_a} \; ,
\end{equation}
acting on the factor of $H_q$ associated with $l_a$
(tensored with the identity operator on the other factors).

Expressed in the Schr\"odinger Picture the class operator is
\begin{align}\label{eq:classoperatoragain}
C(\Phi^n) &=  U(n) P_{2n}({\Phi^n_{2n}})\;
P_{2n-1}({\Phi^n_{2n-1}})\; R_n \dots\nonumber\\
{}& \quad \quad \dots P_4({\Phi^n_{4}})\;
P_3({\Phi^n_{3}}) \; R_2 \;P_2({\Phi^n_{2}})\;
P_1({\Phi^n_{1}}) \;R_1\;,
\end{align}
which might be summarised by the
slogan ``evolve, project, evolve, project...''

We define a useful vector valued amplitude for the finite history
$\Phi^n$ by applying its class operator to the initial state,
\begin{equation}\label{eq:unitaryamp}
\ket{\Phi^n} \equiv C(\Phi^n) \ket{\psi_0}\; .
\end{equation}
This vector is sometimes referred to in the literature as a ``branch'' 
\cite{Hartle:1992as}.

The decoherence functional, $D_q$, is
defined on cylinder sets by the standard expression \cite{Hartle:1992as}
\begin{equation}\label{eq:Dquantum}
D_q(Cyl(\Phi^n)\,; Cyl(\m{\Phi}{}^m))
\equiv \braket{\Phi^n}{{}\,\m{\Phi}{}^m}\,.
\end{equation}
The decoherence functional is defined on the whole event
algebra, $\EA_q$, by additivity since all events are finite disjoint
unions of cylinder sets. Although we have used the natural labelling
 that we chose for the vertices at the beginning, the decoherence
functional thus constructed is independent of the chosen order and
depends only on the vertices' causal order because the projectors
and unitary evolution operators for spacelike separated vertices and
links commute \cite{Dowker:2002wm}.

Note that the properties of the projectors ensure that the formula
(\ref{eq:Dquantum}) for the decoherence functional is consistent with
the condition of additivity when one cylinder set is a disjoint
union of other cylinder sets. For example, $Cyl(\Phi^n)$ is a disjoint
union of all events $Cyl(\Phi^{n+1})$  such that $\Phi^{n+1}$ agrees
with $\Phi^n$ on the first $2n$ links and the decoherence functional
of $Cyl(\Phi^n)$ (with any other event $B$) is indeed given as a sum:
\begin{equation}
D_q(Cyl(\Phi^n)\,;B) =
\mathop{\sum_{\Phi^{n+1}\ {\text s.t.}} }_{\Phi^{n+1}|_n = \Phi^n}
D_q(Cyl(\Phi^{n+1})\,; B)\, ,
\end{equation}
where the sum is over all four field configurations on the
 first $2(n+1)$ links which
agree with $\Phi^n$ on the first $2n$ links.

If the initial
state is a mixed state then the decoherence functional is
a convex combination of pure state decoherence functionals.

This decoherence functional gives a level $2$\ measure, $\mu_q$, 
on $\EA_q$ (see section \ref{sec:QMT}).

\subsection{The collapse model with the Bell ontology}

The above unitary quantum field theory inspired a collapse model
field theory \cite{Dowker:2002wm} which, with the Bell ontology, can be
understood as a level 1 (classical) measure theory in the
Sorkin hierarchy (see section \ref{sec:QMT}) as follows.

The space, $\Omega_c$ of all possible histories/formal trajectories is
an identical copy of that for the quantum field theory, namely
the set of all field configurations on the semi-infinite lattice
to the future of $\sigma_0$.
We will refer to field configurations in
$\Omega_c$ as $\alpha$ in order to distinguish
them from the elements of $\Omega_q$ which we refer to
(as above) as $\Phi$. The event algebra $\EA_c$
consists of finite unions of cylinder sets of elements of $\Omega_c$
and so is isomorphic to $\EA_q$.

The dynamics of the collapse model
is given by a classical (level 1) measure.
Since a level 1 measure is also
level 2 -- each level of the hierarchy includes the
levels below it -- a classical measure can also be given in terms of
a decoherence functional
and in this case the decoherence
functional, $D_c$ is given as follows.

Let $\alpha^n$ be a field configuration on the first $2n$ links.
Define a vector valued amplitude $\ket{\alpha^n}\in H_q$ for each cylinder
set $Cyl(\alpha^n)$:
\begin{equation}
\ket{\alpha^n} \equiv J_{2n}({\alpha^n_{2n}})\;
J_{2n-1}({\alpha^n_{2n-1}})\; R_{n} \dots R_2\; J_2({\alpha^n_{2}})\;
J_1({\alpha^n_{1}})\; R_1 \ket{\psi_0}\, ,
\end{equation}
where $\ket{\psi_0}$ is the initial state on $\sigma_0$ and
$J_a({\alpha^n_a})$ is the Kraus operator implementing a
``partial collapse'' onto the eigenspace corresponding to the value
of $\alpha^n$  at link $l_a$. More precisely,
\begin{align}
J_a(0) &=\frac{1}{\sqrt{1+X^2}}(\ket{0}\bra{0} + X \ket{1}\bra{1}) \label{eqn:J(0)}
\\J_a(1) &=\frac{1}{\sqrt{1+X^2}}\left(X \ket{0} \bra{0} + \ket{1}\bra{1}\right)\, \label{eqn:J(1)}
\end{align}
(where $0\le X\le 1$) acting on the factor of
$H_q$ associated with link $l_a$
(tensored with the identity operator for the other factors).

Then the decoherence functional, $D_c$ is
defined on cylinder sets by
\begin{equation}\label{eq:Dclassical}
D_c(Cyl(\alpha^n)\,; Cyl(\m{\alpha}{}^n)) \equiv
\braket{\alpha^n}{{}\,\m{\alpha}{}^n}\delta_{\alpha^n\, \m{\alpha}{}^n}\, ,
\end{equation}
where $\delta_{\alpha^n\, \m{\alpha}{}^n}$ is a Kronecker delta which is
1 if the two field configurations are identical on all $2n$ links
and zero otherwise.

 The decoherence functional is then extended to the whole
event algebra, $\EA_q$ by additivity since all events are finite
disjoint unions of cylinder sets. In particular, if $m>n$, the
cylinder set $Cyl(\Phi^n)$ with time extent $n$ is a disjoint union
of cylinder sets with time extent $m$, and so it suffices to define $D_q$
as above for cylinder sets of the same time extent:
$D_c(Cyl(\alpha^n)\,; Cyl(\m{\alpha}{}^m))$ is given by additivity.

Again, the decoherence functional thus constructed is independent of
the chosen natural labelling and depends only on the vertices' causal order
because of spacelike commutativity of the evolution operators and
Kraus operators.

$D_c$ is well-defined, in particular the additivity condition is
consistent with the definition (\ref{eq:Dclassical}). For example,
consider
\begin{equation*}
D_c(Cyl(\alpha^n)\,;Cyl(\alpha^n))\,.
\end{equation*}
The event $Cyl(\alpha^n)$ is a disjoint union of all events
$Cyl(\alpha^{n+1})$ for which $\alpha^{n+1}$ agrees with $\alpha^n$
on the first $2n$ links and indeed we have:
\begin{equation}
D_c(Cyl(\alpha^n)\,; Cyl(\alpha^n)) = \mathop{\sum_{\alpha^{n+1}\
{\text s.t.}}}_{\alpha^{n+1}|_n = \alpha^n}\;
\mathop{\sum_{\m{\alpha}{}^{n+1}\ {\text s.t.}}}_{\m{\alpha}{}^{n+1}|_n
=
 \alpha^n}
D_c(Cyl(\alpha^{n+1})\,; Cyl(\m{\alpha}{}^{n+1}))\, .
\end{equation}
In verifying this, the crucial property is that
of the Kraus operators: $J_0^2 + J_1^2 = 1$ and the fact that distinct histories
have no interference, as expressed by the Kronecker delta. Note that
without the Kronecker delta, equation (\ref{eq:Dclassical}) would not
be a consistent definition of a decoherence functional satisfying
additivity.

This decoherence functional is level 1 (classical): it satisfies
\begin{equation}
D_c(Y\,; Z) = D_c(Y\cap Z\,; Y\cap Z)
\end{equation}
and this implies the Kolmogorov sum rule is satisfied by the
measure $\mu_c$ defined by $\mu_c(Y) \equiv D_c(Y\,;Y)$.
Being a level 1 measure, $\mu_c$
has a
familiar interpretation as a probability measure. Indeed the
measure $\mu_c$ defined on the cylinder sets is enough, via the
standard methods of measure theory, to define a unique probability
measure on the whole sigma algebra generated by the cylinder sets.
There is, as yet, no  analogous result for a quantal measure such
as $\mu_q$. Moreover,
there is, as yet, no consensus on how to {\textit {interpret}} a quantum
measure theory. We will not address this important
question here but refer to
\cite{Sorkin:2006wq,Sorkin:2007,Dowker:2007zz} for a new proposal for an
interpretation of quantum mechanics within the framework of
quantum measure theory.

\subsection{Quantum and Classical}

In every collapse model there is a coupling
between classical stochastic variables and a quantum state. How is
this classical-quantum coupling manifested in the generalised
measure theory form of the lattice collapse model just given? We now
show that there is indeed a quantum measure lurking within
and we will expose the nature of the interaction of the quantal variables
with the classical
 variables.

Consider a space of histories $\Omega_{qc}$ which is a direct product of
the two spaces introduced above, $\Omega_{qc}= \Omega_q \times \Omega_c$,
so that elements of $\Omega_{qc}$ are pairs of lattice field configurations,
$(\Phi, \alpha)$. We will refer to the elements of
$\Omega_q$ as {\textit{quantum histories/fields}}
and those of $\Omega_c$ as {\textit {classical histories/fields}}.
The event algebra $\EA_{qc}$ is the ring of sets
generated by the cylinder sets, $Cyl(\Phi^n, \alpha^n)$, where the
cylinder set contains all pairs $(\Phi, \alpha)$ such that $\Phi$
coincides with $\Phi^n$
 and $\alpha$ coincides with $\alpha^n$ on the first $2n$ links.

We now construct a decoherence functional on $\EA_{qc}$ by taking the
unitary decoherence functional, $D_q$ on $\EA_q$, defined above and
``tying'' the classical histories to the quantum histories by
suppressing the decoherence functional by an amount that depends on
how much the classical and quantum field configurations differ. The
more they differ, the greater the suppression. In detail, define
$D_{qc}$ on $\EA_{qc}$ by first defining it on the
cylinder sets:
\begin{align}\label{eq:Dcoupled}
D_{qc}(Cyl(\Phi^n,\alpha^n)\,;\, &Cyl(\m{\Phi}{}^n,\m{\alpha}{}^n)) \equiv
\nonumber \\
{}&D_q(Cyl(\Phi^{n})\,; Cyl( \m{\Phi}{}^{n}))\; \frac{ X^{d(\Phi^n,\alpha^n)+
d(\m{\Phi}{}^n,\m{\alpha}{}^n)} } { (1+X^2)^{2n}} \;
\delta_{\alpha^n\,\m{\alpha}{}^n}
\end{align}
where $0\le X \le 1$ and $d(\Phi^n,\alpha^n)$ is equal to the number of links
on which $\Phi^n$ and $\alpha^n$ differ. As usual it suffices to
define $D_{qc}$ for arguments which have the same time extent, $n$,
because a cylinder set with time extent $m < n$ is a finite disjoint
union of cylinder sets with time extent $n$. $D_{qc}$ is extended to the
full event algebra by additivity.

Checking that the definition (\ref{eq:Dcoupled}) of $D_{qc}$
on the cylinder sets is consistent with the
property of additivity follows the same steps
as for $D_c$ and $D_q$. $D_{qc}$ is
level 2 in the Sorkin hierarchy, although it is clearly
classical on $\Omega_c$.

We now prove some lemmas regarding $D_{qc}$ which
lay bare the structure of our collapse model
of a lattice field in histories
form.

\begin{lemma}
Let $(\Omega_q, \EA_q, D_q)$,
 $(\Omega_c, \EA_c, D_c)$ and  $(\Omega_{qc}, \EA_{qc}, D_{qc})$
be defined as above for the lattice field theory.
Then the decoherence functional for the collapse model, $D_c$
is equal to $D_{qc}$ coarse grained over $\Omega_q$:
\begin{equation}
D_c(A\,;\m{A}) = D_{qc}(\Omega_q\times A\,; \Omega_q\times \m{A}) \ \ \forall
A, \m{A} \in \EA_c\;.
\end{equation}
\end{lemma}
\begin{proof}

It suffices to prove that
\begin{equation}
D_c(Cyl(\alpha^n)\,;Cyl(\m{\alpha}{}^n)) =
\sum_{\Phi^n, \m{\Phi}{}^n}
D_{qc}(Cyl(\Phi^n,\alpha^n)\,;Cyl(\m{\Phi}{}^n,\m{\alpha}{}^n))\;,
\end{equation}
where the double sum is over all field configurations,
$\Phi^n$ and $\m{\Phi}{}^n$,
on the first $2n$ links.
The result follows by additivity because
\begin{equation}
\bigcup_{\Phi^n}Cyl(\Phi^n,\alpha^n) = \Omega_q \times Cyl(\alpha^n)\;.
\end{equation}

Recall
the definition of $D_c$,
\begin{equation*}
D_c(Cyl(\alpha^n)\,; Cyl(\m{\alpha}{}^n)) =
\braket{\alpha^n}{{}\,\m{\alpha}{}^n}\delta_{\alpha^n\, \m{\alpha}{}^n}\, ,
\end{equation*}
where
\begin{equation*}
\ket{\alpha^n} =  J_{2n}({\alpha^n_{2n}})\,
J_{2n-1}({\alpha^n_{2n-1}}) \, R_n \dots R_2 \,J_2({\alpha^n_{2}})\,
J_1({\alpha^n_{1}}) \,R_1 \ket{\psi_0}\, .
\end{equation*}
 Each jump operator $J_a(\alpha^n_a)$
is a linear combination of the two projection
operators $P_a(1)= \ket{1}\bra{1}$ and $P_a(0)= \ket{0}\bra{0}$
on link $l_a$ (see equations \ref{eqn:J(0)} and \ref{eqn:J(1)}). Substituting in this linear combination of projectors for each
 $J_a(\alpha^n_a)$ and expanding out,
 we see that the ket becomes a sum of
$2^{2n}$ terms, one for each possible field configuration
-- call it $\Phi^n$ -- on the $2n$ links. Each such term
is precisely the vector valued amplitude
$\ket{\Phi^n}$ (\ref{eq:unitaryamp}) and
each term is weighted by a factor
\begin{equation*}
\frac{X^{d(\alpha^n, \Phi^n)}}{(1+X^2)^n}
\end{equation*}
from which the result follows.
\end{proof}

The next lemma shows that
if we coarse grain $D_{qc}$ over the classical histories
instead, we find a
quantum theory exhibiting the
symptoms of environmental decoherence.
\begin{lemma}\label{lemma:decoh}
Define a decoherence functional $\widetilde{D}_q$ on $\Omega_q$
by
\begin{equation}
\widetilde{D}_q(F\,; \m{F})  \equiv
D_{qc}(F\times \Omega_c\,; \m{F}\times \Omega_c)\ \ \forall F, \m{F} \in \EA_q \;.
\end{equation}
Then
\begin{equation}
\widetilde{D}_q(Cyl(\Phi^n)\,; Cyl(\m{\Phi}{}^n))=
\left(\frac{2X}{1+X^2}\right)^{d(\Phi^n, \m{\Phi}{}^n)}
D_q(Cyl(\Phi^n)\,; Cyl(\m{\Phi}{}^n))\;.
\end{equation}
\end{lemma}

We leave the proof to the appendix.
Note that the factor suppresses off-diagonal terms in the
decoherence functional and so looks as if it is the result
of environmental decoherence.

\subsection{Equivalence to a model with environment}

The system described by decoherence functional $D_{qc}$ on
the joint space $\Omega_{qc}$ was not derived from any
physical consideration but simply invented as a way to unravel
the decoherence functional of the collapse model. However,
once obtained, the urge to coarse
grain $D_{qc}$ over the classical histories
is irresistible and
the ``approximately diagonal'' form of the resulting
decoherence functional, $\widetilde{D}_q$ on $\Omega_q$ suggests it
can be interpreted as having arisen from coupling to
an environment.

Indeed, the mathematics of collapse models and of
open quantum systems that result from coarse graining
over an ignored environment are known to be closely related
and so it is of no surprise to discover that our
current model can be understood in this way.
Indeed, the classical histories in the collapse model
can simply be reinterpreted as
histories of an environment consisting of variables, one
per link, that interact impulsively with the field there, and then
have no further dynamics.

Let the quantum lattice field,
$\Phi$, interact with
a collection of environment variables, one for each link, taking
values $0$ or $1$.
The space of histories for the whole system
is $\Omega_{qe} \equiv \Omega_q \times \Omega_e$,
where the space of environment
histories, $\Omega_e$, is yet another copy of the same space of
 $\{0,1\}$-field configurations on the semi-infinite lattice. We denote
an element of $\Omega_e$ by $E$, an environment configuration on the
first $2n$ links by $E^n$, the corresponding cylinder set
by $Cyl(E^n)$, and the value of the environment
variable on link $a$ by $E^n_a$.

In the standard state vector language, the Hilbert
space of the whole system of field, $\Phi$,
and the environment is  $H_{qe} \equiv H_q \otimes H_e$
where the environment
Hilbert space, $H_e$, is an infinite tensor product of qubit Hilbert spaces,
$H_{e_a}$, $a = 1,2,\dots$,
one for each link $l_a$ on the lattice to the future of $\sigma_0$.

\begin{lemma}
There is a unitary dynamics of this system such that the unitary
decoherence functional which encodes it, $D_{qe}$,
is equal to $D_{qc}$ if the
environment histories are identified with the classical
histories.
\end{lemma}

\begin{proof}
The proof is by construction of such a dynamics. We add,
to the unitary dynamics of the field $\Phi$,
a one-time interaction
between $\Phi$ and the environment variable on each link
which establishes a partial correlation
between them. Since each environment state lives on
exactly one link, it interacts only once and is
then fixed, which means that
the decoherence functional is diagonal on
the environment histories.

We begin with the space of histories $\Omega_{qe} = \Omega_q \times \Omega_e$
and the Hilbert space $H_{qe} =  H_q \otimes H_e$
where $H_e = \otimes_{a=1}^\infty H_{e_a}$ and each $H_{e_a}$ is a
qubit space.

 The initial state is a
tensor product:
\begin{equation}\label{eq:Psi0}
\ket{\Psi_0} = \ket{\psi_0}_q \otimes_{a =1}^\infty
\ket{0}_{e_a}
\end{equation}
where $ \ket{\psi_0}_q \in H_q$ is the same initial state
for the field $\Phi$ as we had before.

After each elementary unitary evolution $R_i$ is
applied over vertex $i$, two unitary ``partial measurement'' operators
$U_{2i-1}$ and $U_{2i}$ -- to be defined --
 are applied to the Hilbert spaces
associated with
the outgoing links $l_{2i-1}$ and $l_{2i}$, respectively.

Consider a single link, $l_a$. The factor of the total Hilbert space
associated with $l_a$ is
the four-dimensional tensor product of the qubit space, $H_{q_a}$,
of the $\Phi$ states on $l_a$ and the qubit space $H_{e_a}$.
In the field representation, the basis of
this link Hilbert space is $\{\ket{0}_{q_a}\ket{0}_{e_a},
\ket{1}_{q_a}\ket{0}_{e_a},\ket{0}_{q_a}\ket{1}_{e_a},\ket{1}_{q_a}\ket{1}_{e_a}\}$.

The unitary partial measurement operator $U_a$ is defined by
its action on this basis:
\begin{align}
U_a\ket{0}_q\ket{0}_e &= \frac{1}{\sqrt{1+X^2}}\; \ket{0}_q \big(\ket{0}_e + X \ket{1}_e\big)\nonumber\\
U_a\ket{1}_q\ket{0}_e &= \frac{1}{\sqrt{1+X^2}}\; \ket{1}_q \big(X\ket{0}_e +  \ket{1}_e\big)\nonumber\\
U_a\ket{0}_q\ket{1}_e &= \frac{1}{\sqrt{1+X^2}}\; \ket{0}_q \big(X\ket{0}_e - \ket{1}_e\big)\nonumber\\
U_a\ket{1}_q\ket{1}_e &= \frac{1}{\sqrt{1+X^2}}\; \ket{1}_q \big(\ket{0}_e - X \ket{1}_e\big)\label{eq:defU}\;,
\end{align}
where $0\le X\le 1$ and
we have suppressed the $a$ label on all the kets.
$U_a$ acts as the identity on all other factors in
the tensor product Hilbert space for the system.

The action of $U_a$ is to leave $\Phi$ eigenstates alone and put the
initial $\ket{0}_e$ environment state into a superposition of $\ket{0}_e$
and $\ket{1}_e$, so that the environment eigenstate that is correlated
with the $\Phi$ eigenstate is relatively enhanced by a factor $X^{-1}$.

For each cylinder set $Cyl(\Phi^n, E^n)$
we define a vector
valued amplitude, $\ket{\Phi^n, E^n}_{qe} \in H_{qe}$
by evolving the state over each vertex, applying the unitary partial
measurements on the outgoing links and projecting onto the
values of $\Phi^N$ and $E^N$ on the links:
\begin{align}
\ket{\,\Phi^n, E^n}_{qe} \equiv \,&
Q_{2n}(E^n_{2n})\, P_{2n}(\Phi^n_{2n})
\, Q_{2n-1}(E^n_{2n-1})\, P_{2n-1}(\Phi^n_{2n-1})\nonumber\\
{}&\ \ \ \ U_{2n}\, U_{2n-1}\, R_n \dots \nonumber\\
{}&\ \ \ \ \ \ \ \ \dots Q_2(E^n_2)\, P_2(\Phi^n_{2})\,
Q_1(E^n_1) \,P_1(\Phi^n_{1})\nonumber\\
{}&\ \ \ \ \ \ \ \ \ \ \ \ \ \ \ \ \ \ U_2\, U_1 \,R_1 \ket{\Psi_0}\, ,
\end{align}
where $\ket{\Psi_0}$ is defined in (\ref{eq:Psi0}),
$P_a({\Phi^n_a})$ is the projection operator onto the eigenspace
corresponding to the value of $\Phi^n$  at link $l_a$
and $Q_a(E^n_a)$ is the projection operator onto the
eigenspace corresponding to the value of $E^n$ at link
$l_a$. $P_a({\Phi^n_a})$ is only non-trivial on the
factor in $H_q$ associated with link $l_a$ and
$Q_a(E^n_a)$ is only non-trivial on the factor in $H_e$
associated with link $l_a$. As a consequence,
the $P$ projectors and $Q$ projectors
commute.

The initial state is a product, each $U_a$ leaves $\Phi$-eigenstates
alone and the $Q$ projectors act only on the environment states.
We claim that therefore $\ket{\Phi^n, E^n}_{qe}$ is a product,
\begin{equation}
\ket{\Phi^n, E^n}_{qe} = \ket{\Phi^n}_q \ket{E^n}_e\,,
\end{equation}
where $\ket{\Phi^n}_q \in H_q$ is the vector valued amplitude (\ref{eq:unitaryamp})
for the plain vanilla unitary field theory
and
\begin{equation}\label{eq:ketE}
\ket{E^n}_{e} = \frac{X^{d(\Phi^n, E^n)}}{(1+X^2)^n}\;
\ket{E^n_1}_{e_1} \ket{E^n_2}_{e_2}\dots \ket{E^n_{2n}}_{e_{2n}}
\end{equation}
where we have left off the factors of $\ket{0}$ for all the
infinitely many links
to the future of $\sigma_n$, which play no role.

The proof of this claim is given in the appendix.

The decoherence functional, $D_{qe}$, for the total system
is given by
\begin{align}
D_{qe}(Cyl(\Phi^n,E^n)\,;Cyl(\m{\Phi}{}^n,\m{E}{}^n)) &\equiv
\braket{\Phi^n,E^n}{\m{\Phi}{}^n,\m{E}{}^n} \\
{}& = \braket{\Phi^n}{\m{\Phi}{}^n}_q \braket{E^n}{\m{E}{}^n}_e\,.
\end{align}
Using (\ref{eq:ketE}), we see that the decoherence functional is
zero unless $E^n = \m{E}{}^n$ and we have
\begin{align}
D_{qe}(Cyl(\Phi^n,E^n)\,;\,&Cyl(\m{\Phi}{}^n,\m{E}{}^n)) = \nonumber\\
D_q(Cyl(\Phi^{n})\,; Cyl(\m{\Phi}{}^{n}))\; \frac{X^{d(\Phi^n,E^n)+
d(\m{\Phi}{}^n,\m{E}{}^n)}}{(1+X^2)^{2n}} \;\delta_{E^n\,\m{E}{}^n}\;.
\end{align}

As usual, we only need to define the decoherence functional
for cylinder sets of equal time extent. We see that this is
equal to $D_{qc}$, the decoherence functional of the
collapse model (\ref{eq:Dcoupled}).
\end{proof}

The model is technically unitary and so falls into
the category of ordinary quantum theory, but the classicality of the
environment variables is achieved by the device of postulating an infinite
environment and one-time interactions.

\section{Discussion}
None of the physics we have presented is new.
We have merely provided a novel perspective
on a known model that arises when spacetime
and histories are given a central role.
Di\'osi stressed that both classical variables and quantum state are
present in a collapse model and advocates ascribing reality to them both
\cite{Diosi:2004}.
We have replaced the formalism of
quantum state with quantum histories and
by placing quantum and classical variables on the same
footing in spacetime
we can see more clearly the character of the interaction
between them.

We claim that the structure outlined above for the collapse model
for a lattice field theory, is generic to collapse models.
There is always, more or less hidden in the model,
a  space of histories which is a product of a
space of quantum histories and a space of classical histories,
with a decoherence functional on it.
For example, in the case of the GRW model \cite{Ghirardi:1986mt}
the classical histories
are countable subsets of Galilean spacetime, to the future of some
initial surface, $t=0$. The elements of such a countable subset are
the ``collapse centres'' $(x_i, t_i), i = 1,2, \dots$.
The probability distribution on these
classical histories is given by a classical
decoherence functional $D_c$, which is, essentially, set out in
\cite{Kent:1998bc}.
In order to follow the steps taken in this paper of
unravelling $D_c$ into $D_{qc}$, the positive operators, Gaussians,
that correspond to the classical events are
expressed as
integrals of projection operators and the evolution between
collapses expressed using the Dirac-Feynman propagator as
a sum over the histories.
The quantum histories, then, are precisely the
histories summed over in the Dirac-Feynman path integral: all
continuous real functions $\gamma: [0,\infty] \rightarrow \RR$.

The continuum limit of the GRW model is the
continuous spontaneous localisation model for a single particle
\cite{Diosi:1988a,Diosi:1988b} and this too can be cast into the generic
form as can be seen
from the formulation of the model in terms of
a ``restricted propagator'' as described in
references \cite{Mensky:1979, Mensky:1994, Diosi:1995}.
Although the analysis in these references uses phase space path
integrals, if it is the position operator whose eigenstates are collapsed onto, as is the case for the continuum limit of GRW, the path
integrals can be transformed into configuration space path integrals.
In this case, the quantum histories are again the continuous paths
that contribute to the Dirac-Feynman sum-over-histories,
but the classical histories are very noisy, and not continuous
paths at all.

Note that in the lattice field theory the spaces of classical and quantum
histories in this case are isomorphic, whereas in the GRW model and its
continuum limit the quantum and classical histories are very different.
In all cases, however, it is the quantum histories that bear all the consequence
of dynamical law encoded in a local spacetime
action, whereas the classical histories
are simply dragged along by being tied to the quantum histories.

This state of affairs is illuminated further by considering
coupling together two separate collapse models X and Y. Each model will
contain both quantum and classical histories and the
coupling between X and Y will be achieved by
an appropriate term in the action involving the quantum histories alone.
It is the quantum histories of X which directly touch the quantum histories of
Y. The classical variables of X only react to the classical
variables of Y because they are restricted to be close to the
quantum variables which interact with the quantum variables
of Y to which the classical variables of Y must, in their turn, be close.

The present authors believe,
with Hartle, Sorkin and others, that a spacetime approach to quantum
mechanics will be
essential to progress in quantum gravity and for this reason
spacetime approaches should be  carefully studied.
Two important reasons for
pursuing collapse models with the Bell ontology are that
the models are already in spacetime form and
the stochasticity involved is completely classical so all the familiar machinery
of stochastic processes can be brought to bear:
the stochasticity of collapse models causes no more
interpretational difficulty than does the randomness of Brownian
motion.
The theory concerns the classical variables only
and the quantum histories are relegated to some sort of
auxiliary, hidden status, despite the fact that
the dynamics of the model is most easily
described in terms of these quantum histories.
In order to pursue this direction, therefore, one must
pay the price of
ignoring the quantum histories as far as the ontology is
concerned: ``Pay no attention to that man behind the
curtain'' \cite{Oz:1939}.

On the other hand, if the quantum histories are kept in the theory
to be treated on the same footing, a priori, as the
classical histories, then
the question of the physical meaning of the
quantum measure on them has to be wrestled with:
what {\it is} the ontology in a quantum measure theory?
But if {\textit {this}} thorny problem is to be tackled, then
one might start by trying to address it in the
case of unitary quantum mechanics in the
first instance. It may be that an interpretation of the
quantum measure can be discovered that,
 by itself, provides
a solution to the interpretational problems of quantum mechanics,
while yet maintaining unitary dynamics
and without need of new quantum-classical couplings.

\section{Acknowledgments}
We thank Rafael Sorkin for invaluable discussions throughout this work.
We thank Isabelle Herbauts for useful discussions.
YG is supported by a PPARC
studentship. FD is supported in part by
the EC Marie Curie Research and Training Network,
Random Geometry and Random Matrices MRTN-CT-2004-005616
(ENRAGE) and by Royal Society International Joint Project 2006/R2.

\appendix
\section{Appendix}

\begin{proof}\;  \textit{{Of Lemma}} \ref{lemma:decoh}

Recall the definition of $D_{qc}$:
\begin{align*}
D_{qc}(Cyl(\Phi^n,\alpha^n)\,;\,& Cyl(\m{\Phi}{}^n,\m{\alpha}{}^n)) =\\
{}&D_q(Cyl(\Phi^{n})\,; Cyl(\m{\Phi}{}^{n}))\; \frac{X^{d(\Phi^n,\alpha^n)+
d(\m{\Phi}{}^n,\m{\alpha}{}^n)}}{(1+X^2)^{2n}} \;\delta(\alpha^n,\m{\alpha}{}^n)\;.
\end{align*}
When the sum is taken over all $\alpha^n$ and $\m{\alpha}{}^n$, field
configurations on the first $2n$ vertices,
 it results in
 \begin{align}
D_{qc}(Cyl(\Phi^n)\times \Omega_c\,;\,& Cyl(\m{\Phi}{}^n)\times \Omega_c) =
\nonumber \\
{}& \frac{1}{(1+X^2)^{2n}}\;  D_q(Cyl(\Phi^{n})\,; Cyl(\m{\Phi}{}^{n}))
\sum_{\alpha^n}  X^{d(\Phi^n,\alpha^n)+
d(\m{\Phi}{}^n,\alpha^n)}\;.
\end{align}

Let $d(\Phi^n, \m{\Phi}{}^n) = m$, which is the number of links on
which the values of the two fields differ. For the duration
of this proof only, we relabel the links on which the two
fields differ $l_1,l_2,\dots l_m$ and the rest,
on which the fields agree, are labelled $l_{m+1},\dots l_{2n}$.
Consider the exponent ${d(\Phi^n,\alpha^n)+
d(\m{\Phi}{}^n,\alpha^n)}$. The first $m$ links contribute $m$ to
the exponent whatever $\alpha^n$ is,
because for each link, $\alpha^n$ will agree with
exactly one of $\Phi^n$ and $\m{\Phi}{}^n$. Therefore
\begin{equation}
{d(\Phi^n,\alpha^n)+
d(\m{\Phi}{}^n,\alpha^n)} =
m + 2\tilde{d}(\alpha^n, \Phi^n)\,,
\end{equation}
where $\tilde{d}$ is the number of the last $2n -m$ links
on which $\alpha^n$ and $\Phi^n$ differ.

The sum over $\alpha^n$
can be expressed as a multiple
sum over the values of the $\alpha$ variable on each
link in turn. We first do the sum over the values on the $m$ links
on which $\Phi^n$ and $\m{\Phi}{}^n$ differ. The summand does not depend
on the values on those links and so that gives a factor of $2^m$
 \begin{equation}
\sum_{\alpha^n} X^{d(\Phi^n,\alpha^n)+
d(\m{\Phi}{}^n,\alpha^n)}
=  2^m X^m \sum_{\alpha^n_{m+1}} \dots \sum_{\alpha^n_{2n}}
X^{2\tilde{d}(\alpha^n, \Phi^n)}\;.
\end{equation}
The remaining sum is over all $\alpha$ configurations on the last
$2n-m$ links. There is one such configuration that agrees with
$\Phi^n$ on all $2n-m$ links, $\binom{2n-m}{1}$ configurations
that differ from $\Phi^n$ on one link,
$\binom{2n-m}{2}$  that differ from $\Phi^n$ on two links, {\textit {etc}}.
The remaining sum therefore gives $(1+X^2)^{2n-m}$ and we have
 \begin{equation}
\sum_{\alpha^n} X^{d(\Phi^n,\alpha^n)+
d(\m{\Phi}{}^n,\alpha^n)}
=  2^m X^m (1+X^2)^{2n-m}\,,
\end{equation}
and hence the result.

\end{proof}

\begin{claim}

\begin{equation}
\ket{\,\Phi^n, E^n}_{qe} =
\frac{X^{d(\Phi^n, E^n)}}{(1+X^2)^n}\; \ket{\Phi^n}_q
\ket{E^n_1}_{e_1} \ket{E^n_2}_{e_2}\dots
\ket{E^n_{2n}}_{e_{2n}}\otimes_{a=2n+1}^\infty \ket{0}_{e_a} \,,
\end{equation}
where $\ket{\Phi^n}_q$ is given by (3.1).

This is the claim in lemma 3.
\end{claim}
\begin{proof}

We use induction. It is trivially true for $n=0$.

We assume it is true for $n$. Let $\Phi^{n+1}|_n = \Phi^n$ and
$E^{n+1}|_n = E^n$. Then
\begin{align}
\ket{\,\Phi^{n+1}, E^{n+1}}_{qe} ={}&\;
Q_{2n+2}(E^{n+1}_{2n+2})\, P_{2n+2}(\Phi^{n+1}_{2n+2})
\, Q_{2n+1}(E^{n+1}_{2n+1})\, P_{2n+1}(\Phi^{n+1}_{2n+1})\nonumber\\
{}&\ \ \ \ U_{2n+2}\, U_{2n+1}\, R_{n+1}\ket{\Phi^n,E^n}_{qe}\,.
\end{align}

The $P$ projectors commute with the $Q$ projectors. The $P_{a}$ projectors
also commute with the partial measurement operators $U_{a}$ as can be seen from
the definition of $U$ (\ref{eq:defU}). So we have
\begin{align}
\ket{\,\Phi^{n+1}, E^{n+1}}_{qe} = \,&\,\frac{X^{d(\Phi^n, E^n)}}{(1+X^2)^n}\;
Q_{2n+2}(E^{n+1}_{2n+2})\, Q_{2n+1}(E^{n+1}_{2n+1})\,  U_{2n+2}\, U_{2n+1}
\nonumber\\
{}&\ \ \ \ \ \ \ \left[\,P_{2n+2}(\Phi^{n+1}_{2n+2})\,P_{2n+1}(\Phi^{n+1}_{2n+1})\,R_{n+1}
 \ket{\Phi^n}_q \,\right]\nonumber \\
{}&\ \ \ \ \ \ \ \ \ \ \ \ \ \ket{E^n_1}_{e_1} \dots
\ket{E^n_{2n}}_{e_{2n}} \ket{0}_{e_{2n+1}}\ket{0}_{e_{2n+2}}
\otimes_{a=2n+3}^\infty \ket{0}_{e_a} \,.
\end{align}

The factor in square brackets is $\ket{\Phi^{n+1}}_q \in H_q$ and is
unchanged by the $U$'s because it is an eigenstate of the field $\Phi$
on the links $l_{2n+1}$ and $l_{2n+1}$. The same factor is also
unchanged by the $Q$'s
which only act on the
environment states.
$U_{2n+1}$ turns $\ket{0}_{e_{2n+1}}$ into a linear combination
of $\ket{0}_{e_{2n+1}}$ and $\ket{1}_{e_{2n+1}}$, enhancing the
term which is correlated to the value $\Phi^{n+1}_{2n+1}$.
Similarly for $U_{2n+2}$. Finally $Q_{2n+1}(E^{n+1}_{2n+1})$
projects onto the state $\ket{E^{n+1}_{2n+1}}_{e_{2n+1}}$ and
similarly for $Q_{2n+2}(E^{n+1}_{2n+2})$ with the result
\begin{equation}
\ket{\,\Phi^{n+1}, E^{n+1}}_{qe} = \frac{X^{d(\Phi^n, E^n)}}{(1+X^2)^n}\;
\frac{ X^{ 2 - \delta(\Phi^{n+1}_{2n+2}, E^{n+1}_{2n+2})-
\delta(\Phi^{n+1}_{2n+1}, E^{n+1}_{2n+1}) }}{(1+X^2)} \,
\ket{\Phi^{n+1}}_q \ket{E^{n+1}}_e \,.
\end{equation}
The $\delta$'s in the exponent of $X$
are Kronecker deltas and combining the factors of
$X$ gives the result.

\end{proof}
\bibliography{../Bibliography/refs} \bibliographystyle{../Bibliography/JHEP}

\providecommand{\href}[2]{#2}\begingroup\raggedright\begin{thebibliography}{10}

\bibitem{Bassi:2003gd}
A.~Bassi and G.~Ghirardi, {\it Dynamical reduction models},  {\em Phys. Rept.}
  {\bf 379} (2003) 257, [\href{http://xxx.lanl.gov/abs/quant-ph/0302164}{{\tt
  quant-ph/0302164}}].

\bibitem{Sorkin:1994dt}
R.~D. Sorkin, {\it Quantum mechanics as quantum measure theory},  {\em Mod.
  Phys. Lett.} {\bf A9} (1994) 3119--3128,
  [\href{http://xxx.lanl.gov/abs/gr-qc/9401003}{{\tt gr-qc/9401003}}].

\bibitem{Bell:1987i}
J.~Bell, {\em Speakable and unspeakable in quantum mechanics}, ch.~18.
\newblock CUP, Cambridge, 1987.

\bibitem{Sorkin:1995nj}
R.~D. Sorkin, {\it Quantum measure theory and its interpretation},  in {\em
  Quantum Classical Correspondence: Proceedings of 4th Drexel Symposium on
  Quantum Nonintegrability, September 8-11 1994, Philadelphia, PA} (D.~Feng and
  B.-L. Hu, eds.), pp.~229--251, International Press, Cambridge, Mass., 1997.
\newblock \href{http://xxx.lanl.gov/abs/gr-qc/9507057}{{\tt gr-qc/9507057}}.

\bibitem{Salgado:1999pu}
R.~B. Salgado, {\it Some identities for the quantum measure and its
  generalizations},  {\em Mod. Phys. Lett.} {\bf A17} (2002) 711--728,
  [\href{http://xxx.lanl.gov/abs/gr-qc/9903015}{{\tt gr-qc/9903015}}].

\bibitem{Sorkin:2006wq}
R.~D. Sorkin, {\it Quantum dynamics without the wave function},  {\em J. Phys.}
  {\bf A40} (2007) 3207--3222,
  [\href{http://xxx.lanl.gov/abs/quant-ph/0610204}{{\tt quant-ph/0610204}}].

\bibitem{Sorkin:2007}
R.~D. Sorkin, {\it An exercise in ``anhomomorphic logic''},  2007.
\newblock to appear in a special volume of {\it Journal of Physics}, edited by
  L. Di\'osi, H-T Elze, and G. Vitiello, [quant-ph/0703276].

\bibitem{Hartle:1989}
J.~B. Hartle, {\it The quantum mechanics of cosmology},  in {\em Quantum
  Cosmology and Baby Universes: Proceedings of the 1989 Jerusalem Winter School
  for Theoretical Physics} (S.~Coleman, J.~B. Hartle, T.~Piran, and
  S.~Weinberg, eds.), World Scientific, Singapore, 1991.

\bibitem{Hartle:1992as}
J.~B. Hartle, {\it Space-time quantum mechanics and the quantum mechanics of
  space-time},  in {\em Proceedings of the Les Houches Summer School on
  Gravitation and Quantizations, Les Houches, France, 6 Jul - 1 Aug 1992}
  (J.~Zinn-Justin and B.~Julia, eds.), North-Holland, 1995.
\newblock \href{http://xxx.lanl.gov/abs/gr-qc/9304006}{{\tt gr-qc/9304006}}.

\bibitem{Dowker:2002wm}
F.~Dowker and J.~Henson, {\it A spontaneous collapse model on a lattice},  {\em
  J. Stat. Phys.} {\bf 115} (2004) 1349,
  [\href{http://xxx.lanl.gov/abs/quant-ph/0209051}{{\tt quant-ph/0209051}}].

\bibitem{Dowker:2004zn}
F.~Dowker and I.~Herbauts, {\it Simulating causal wave-function collapse
  models},  {\em Class. Quant. Grav.} {\bf 21} (2004) 1--17,
  [\href{http://xxx.lanl.gov/abs/quant-ph/0401075}{{\tt quant-ph/0401075}}].

\bibitem{Destri:1987ze}
C.~Destri and H.~J. de~Vega, {\it Light cone lattice approach to fermionic
  theories in 2-d: The massive {T}hirring model},  {\em Nucl. Phys.} {\bf B290}
  (1987) 363.

\bibitem{Dowker:2007zz}
F.~Dowker and Y.~Ghazi-Tabatabai, {\it The {K}ochen-{S}pecker {T}heorem
  {R}evisited in {Q}uantum {M}easure {T}heory},  2007.
\newblock eprint arXiv:0711.0894.

\bibitem{Diosi:2004}
L.~Di\'osi. Talk at `Quantum Theory Without Observers II,' Bielefeld, 2-6
  February 2004, {2004}.

\bibitem{Ghirardi:1986mt}
G.~C. Ghirardi, A.~Rimini, and T.~Weber, {\it A unified dynamics for micro and
  macro systems},  {\em Phys. Rev.} {\bf D34} (1986) 470.

\bibitem{Kent:1998bc}
A.~Kent, {\it Quantum histories},  {\em Phys. Scripta} {\bf T76} (1998) 78--84,
  [\href{http://xxx.lanl.gov/abs/gr-qc/9809026}{{\tt gr-qc/9809026}}].

\bibitem{Diosi:1988a}
L.~Di\'osi, {\it Continuous quantum measurement and {I}t\^o formalism},  {\em
  Phys. Lett.} {\bf A129} (1988) 419--423.

\bibitem{Diosi:1988b}
L.~Di\'osi, {\it Localized solution of a simple nonlinear quantum {L}angevin
  equation},  {\em Phys. Lett.} {\bf A132} (1988) 233--236.

\bibitem{Mensky:1979}
M.~B. Mensky {\em Phys. Rev.} {\bf D20} (1979) 384.

\bibitem{Mensky:1994}
M.~B. Mensky {\em Phys. Lett.} {\bf A196} (1994) 159.

\bibitem{Diosi:1995}
L.~Di\'osi, {\it Selective continuous quantum measurements: Restricted path
  integrals and wave equations},  1995.
\newblock eprint arXiv:quant-ph/9501009.

\bibitem{Oz:1939}
L.~F. Baum, N.~Langley, and V.~Fleming, {\it The {W}izard of {O}z}, Metro-Goldwyn-Mayer, Los Angeles, California, 1939.

\end{thebibliography}\endgroup

\end{document}